# Iterative Hard Thresholding for Compressed Sensing

Thomas Blumensath and Mike E. Davies

**Abstract**

Compressed sensing is a technique to sample *compressible* signals below the Nyquist rate, whilst still allowing near optimal reconstruction of the signal. In this paper we present a theoretical analysis of the iterative hard thresholding algorithm when applied to the compressed sensing recovery problem. We show that the algorithm has the following properties (made more precise in the main text of the paper)

- It gives near-optimal error guarantees.
- It is robust to observation noise.
- It succeeds with a minimum number of observations.
- It can be used with any sampling operator for which the operator and its adjoint can be computed.
- The memory requirement is linear in the problem size.
- Its computational complexity per iteration is of the same order as the application of the measurement operator or its adjoint.
- It requires a fixed number of iterations depending only on the logarithm of a form of signal to noise ratio of the signal.
- Its performance guarantees are uniform in that they only depend on properties of the sampling operator and signal sparsity.

## I. Introduction

For more than fifty years, the Nyquist-Shannon [1] [2] sampling theorem was generally used as the foundation of signal acquisition systems. Using this theory, it was common held belief that signals have to be sampled at twice the signal bandwidth. Whilst this is true for general band-limited signals, this theory does not acount for additional signal structures that might be known a priory. The recently emerging field of compressed sensing, [3], [4], [5], [6] and [7] and the related theory of signals with a finite rate of innovations [8], start from another premise. In compressed sensing, signals are assumed to be sparse in some transform domain. This sparsity constraint significantly reduces the size of the set of possible signals compared to the signal space dimension. This is easily visualised if one imagines the set of all images (say of dimension $256 \times 256$) with pixel values in the range from zero to one. To get an idea of how scant the set of *true* or *interesting* images is in this space, randomly draw images from this set, that is, draw the pixels from uniform distributions. One could spend the rest of once life doing this, without encountering anything remotely resembling an image. In other words, the set of images generally of interest (such as you recent holiday snapshots) occupy a minute subset of all possible images.

This implies that sparse signals have an information content much smaller than that implied by the Nyquist rate. Instead of requiring 65536 numbers to represent the $256 \times 256$ image, if the image has a sparse representation, it is possible to represent the same image using many fewer bits of information. This property of many signals is exploited in compressed sensing. Instead of taking samples at the Nyquist rate, compressed sensing uses linear sampling operators that map the signal into a small (compared to the Nyquist rate) dimensional space. This process is a combination of sampling and compression, hence the name compressed sensing or compressive sampling. Contrary to the Nyquist-Shannon theory, in compressed sensing, reconstruction of the signal is non-linear. One of the important contributions of the seminal work by Candes, Romber, Tao [4], [5], [6] and Donoho [7], was to show that linear programming algorithms can be used under certain conditions on the sampling operator to reconstruct the original signal with high accuracy.

Another set of algorithms, which could be shown to efficiently reconstruct signals from compressed sensing observations are greedy methods. A by now traditional approach is Orthogonal Matching Pursuit [9], which was analysed as a reconstruction algorithm for compressed sensing in [10]. Better theoretical properties were recently 4proven for a regularised Orthogonal Matching Pursuit algorithm [11], [12]. Even more recently, the Subspace Pursuit [13] and the nearly identical Compressive Sampling Matching Pursuit (CoSaMP) [14] algorithms were introduced and analysed for compressed sensing signal reconstruction. Of all of these methods, CoSaMP currently offers the most comprehensive set of theoretic performance guarantees. It works for general sampling operators, is robust against noise and the performance is uniform in that it only depends on a property of the sampling operator and the sparsity of the signal, but not on the size of the non-zero signal coefficients. Furthermore, it requires minimal storage and computations and works with (up to a constant) a minimal number of observations.

In a previous paper [15], iterative hard thresholding algorithms were studied. In particular, their convergence to fixed points of $\ell_0$ regularised (or constrained) cost functions could be proven. In this paper, it is shown that one of these algorithms (termed from now on $IHT_s$) has similar performance guarantees to those of CoSaMP.

*A. Paper Overview*

Section II starts out with a definition of sparse signal models and a statement of the compressed sensing problem. In section III, we then discuss an iterative hard thresholding algorithm. The rest of this paper shows that this algorithm is able to recover, with high accuracy, signals form compressed sensing observations. This result is formally stated in the theorem and corollary in the first subsection of section IV. The rest of section IV is devoted to the proof of this theorem and its corollary. The next section, section V, takes a closer look at a stopping criterion for the algorithm, which guarantees a certain estimation accuracy. The results of this paper are similar to those for the CoSaMP algorithm of [14] and a more detailed comparison is given in section VI.

*B. Notation*

The following notation will be used in this paper. $\mathbf{x}$ is an $M$-dimensional real or complex vector. $\mathbf{y}$ is an $N$ dimensional real or complex vector. $\Phi$ will denote an $M \times N$ real or complex matrix, whose transpose (hermitian transpose) will be denoted by $\Phi^T$. Many of the arguments in this paper will use sub-matrices and sub-vectors. The letters $\Gamma$, $B$ and $\Lambda$ will denote sets of indices that enumerate the columns in $\Phi$ and the elements in the vectors $\mathbf{y}$. Using these sets as subscripts, e.g., $\Phi_\Gamma$ or $\mathbf{y}_\Gamma$, we mean matrixes (or vectors) formed by removing all but those columns (elements) from the matrix (vector) other than those in the set. We also have occasion to refer to quantities in a given iteration. Iterations are counted using $n$ or $k$. For sets, we use the simplified notation $\Gamma^n$ whilst for vectors, the iteration count is given in square brackets $\mathbf{y}^{[n]}$. For the definition of the restricted isometry property used in the research literature, we use the Greek letter $\delta_s$ to denote the restricted isometry constant. However, in this paper, we use a slightly modified definition of the property. The associated restricted isometry constant will be labelled by $\beta_s$ throughout.

The following norms are used repeatedly. $\|\cdot\|_2$ is the Euclidean vector norm or, for matrixes, the operator norm from $\ell_2$ to $\ell_2$. We will also need the vector $\ell_1$ norm $\|\cdot\|_1$. The notation $\|\cdot\|_0$ will denote the number of non-zero elements in a vector. For a general vector $\mathbf{y}$ we use $\mathbf{y}^s$ to be any of the best $s$ term approximations to $\mathbf{y}$. The difference between the two will be $\mathbf{y}_r = \mathbf{y} - \mathbf{y}^s$. The support, that is the index set labeling the non-zero elements in $\mathbf{y}^s$, is defined as $\Gamma^\star = \mathrm{supp}\{\mathbf{y}^s\}$ and similarly, $\Gamma^n = \mathrm{supp}\{\mathbf{y}^{[n]}\}$, where $\mathbf{y}^{[n]}$ is the estimation of $\mathbf{y}$ in iteration $n$. Finally, the set $B^n = \Gamma^\star \bigcup \Gamma^n$ is a superset of the support of the error $\mathbf{r}^{[n]} = \mathbf{y}^s - \mathbf{y}^{[n]}$. Set difference is denoted using $\cdot \setminus \cdot$.

## II. SPARSITY AND COMPRESSED SENSING

*A. Sparse Signal Models*

In this section we formalise the notion of sparsity and sparse signal models. A vector will be called $s$-spare if no more than $s$ of its elements have non-zero values. We will talk of a best $s$-sparse approximation to a vector $\mathbf{y}$ to mean any one of the $s$-sparse vectors $\mathbf{y}^s$ that minimise $\|\mathbf{y}^s - \mathbf{y}\|_2$.

Most signals in the real world are not exactly sparse. Instead, they are often well approximated by an $s$-sparse signal. A good example are images, whose wavelet transform has most of its energy concentrated in relatively few coefficients. To model such behaviour, we use the following notion. Assume the elements in a vector $\mathbf{y}$ are reordered such that $|y_i| \geq |y_{i-1}|$. A signal is called *p-compressible*, if, for some constant $R$, the coefficients satisfy the following property

$$|y_i| \leq R i^{-1/p}, \tag{1}$$

that is, the reordered coefficients decay with a power law. The importance of these *p-compressible* signals is that they can be well approximated using exact sparse signals. Let $\mathbf{y}^s$ be any best $s$-term approximation of a *p-compressible* signal $\mathbf{y}$, then it is easy to show that

$$\|\mathbf{y} - \mathbf{y}^s\|_2 \leq cRs^{1/2-1/p} \tag{2}$$

and

$$\|\mathbf{y} - \mathbf{y}^s\|_1 \leq cRs^{1-1/p}, \tag{3}$$

where $c$ are different constants depending only on $p$, but not on s. Note that the above two terms bound the error in terms of the $\ell_1$ as well as the $\ell_2$ norm. These two norms will appear in the main result derived in this paper and can therefore be used to derive performance guarantees for the recovery of $p$-compressible signals.

### B. Compressed Sensing

We assume signals $\mathbf{y}$ to be representable as real or complex vectors of finite length. Compressed sensing samples such signals, using a linear mapping $\Phi$ into an $M$ dimensional real or complex observation space. In matrix notation, the observed samples $\mathbf{x}$ are

$$\mathbf{x} = \Phi \mathbf{y} + \mathbf{e}. \tag{4}$$

Here, $\Phi$ is the linear sampling operator and $\mathbf{e}$ models possible observation noise due to, for example, sensor noise or quantisation errors in digital systems.

In compressed sensing, two main problems have to be addressed. The first problem, not studied in detail in this paper, is to design measurement systems $\Phi$ that possess certain desirable properties (such as the restricted isometry property of the next subsection), which allow for an efficient estimation of $\mathbf{y}$. The second problem, which is at the focus of this paper, is the study of concrete algorithms for the efficient estimation of $\mathbf{y}$, given only $\mathbf{x}$ and $\Phi$.

### C. The Restricted Isometry Property

The analysis of algorithms for compressed sensing relies heavily on the following property of the observation matrix $\Phi$. A matrix $\hat{\Phi}$ satisfies the Restricted Isometry Condition (RIP) [4] if

$$(1-\delta_s)\|\mathbf{y}\|_2^2 \leq \|\hat{\Phi}\mathbf{y}\|_2^2 \leq (1+\delta_s)\|\mathbf{y}\|_2^2 \tag{5}$$

for all $s$-sparse $\mathbf{y}$. The *restricted isometry constant* $\delta_s$ is defined as the smallest constant for which this property holds for all $s$-sparse vectors $\mathbf{y}$. Instead of using the above restricted isometry property, we will use a re-scaled matrix $\Phi = \frac{\hat{\Phi}}{1+\delta_s}$, which satisfies the following non-symetric isometry property, which is equivalent to the RIP defined above.

$$(1-\beta_s)\|\mathbf{y}\|_2^2 \leq \|\Phi\mathbf{y}\|_2^2 \leq \|\mathbf{y}\|_2^2 \tag{6}$$

for all $s$-sparse $\mathbf{y}$. Now $\beta_s = 1 - \frac{1-\delta_s}{1+\delta_s}$. We will say that for a matrix $\Phi$ the RIP holds for sparsity $s$, if $\beta_s < 1$. Throughout this paper, when referring to the RIP, we mean in general this slightly modified version for which we always use the letter $\beta$. If we need to refer to the standard RIP, for the non-normalised matrix $\hat{\Phi}$, we use the letter $\delta$. This is only required in order to compare our results to others in the literature.

We also need the following properties of the RIP derived in for example [14]. Note that the RIP gives an upper bound on the largest and a lower bound on the s-largest singular values of all sub-matirxes of $\Phi$ with s columns. Therefore, for all index sets $\Gamma$ and all $\Phi$ for which the RIP holds with $s = |\Gamma|$, the following inequalities hold trivially

$$\|\Phi_\Gamma^T \mathbf{x}\|_2 \leq \|\mathbf{x}\|_2, \tag{7}$$

$$(1-\beta_{|\Gamma|})\|\mathbf{y}_\Gamma\|_2 \leq \|\Phi_\Gamma^T \Phi_\Gamma \mathbf{y}_\Gamma\|_2 \leq \|\mathbf{y}_\Gamma\|_2 \tag{8}$$

The following two properties are at the heart of the proof of the main result of this paper and we derive these therefore here.

*Lemma 1:* For all index sets $\Gamma$ and all $\Phi$ for which the RIP holds with $s = |\Gamma|$

$$\|(\mathbf{I} - \Phi_\Gamma^T \Phi_\Gamma)\mathbf{y}_\Gamma\|_2 \leq \beta_s \|\mathbf{y}_\Gamma\|_2. \tag{9}$$

*Proof:* The RIP guarantees that the eigenvalues of the matrix $\Phi_\Gamma^T \Phi_\Gamma$ fall within the interval $[1-\beta_s, 1]$. This can easily be seen to imply that the matrix $\mathbf{I} - \Phi_\Gamma^T \Phi_\Gamma$ has eigenvalues in the interval $[0, \beta_s]$, which proves the lemma. ∎

The next inequality is known and can be found in for example [14].

*Lemma 2:* For two *disjoint* sets $\Gamma$ and $\Lambda$ (i.e. $\Gamma \bigcap \Lambda = \emptyset$) and all $\Phi$ for which the RIP holds with $s = |\Gamma \bigcup \Lambda|$

$$\|\Phi_\Gamma^T \Phi_\Lambda \mathbf{y}_\Lambda\|_2 \leq \beta_s \|\mathbf{y}_\Lambda\|_2. \tag{10}$$

For completeness, we here repeat the proof similar to the one given in [14].

*Proof:* Let $\Omega = \Gamma \bigcup \Lambda$. Because the sets $\Gamma$ and $\Lambda$ are disjoint, the matrix $-\Phi_\Gamma^T \Phi_\Lambda$ is a submatrix of $\mathbf{I} - \Phi_\Omega^T \Phi_\Omega$. By [16, Theorem 7.3.9], the largest singular value of a submatrix is bounded by the largest singular value of the full matrix. In other words, $\|\Phi_\Gamma^T \Phi_\Lambda\|_2 \leq \|\mathbf{I} - \Phi_\Omega^T \Phi_\Omega\|_2$. To bound the operator norm on the right, note that by the RIP, $\Phi_\Omega^T \Phi_\Omega$ has eigenvalues in the interval $[1-\beta_s, 1]$. Therefore $\mathbf{I} - \Phi_\Omega^T \Phi_\Omega$ has eigenvalues in the interval $[0, \beta]$, which proves the lemma. ∎

### D. Designing Measuring Systems

The RIP is a condition on the singular values of sub-matrixes of a matrix $\Phi$. To use the results based on the RIP in practice, one would need to either, be able to calculate the RIP for a given matrix, or be able to design a matrix with a pre-specified RIP. Unfortunately, none of the above is possible in general. Instead, the only known approach to generate a matrix $\Phi$ satisfying the RIP with high probability, is to use random constructions. For example, if the entries in $\Phi$ are drawn independently from certain probability distributions (such as a Gaussian distribution or a Bernoulli distribution), with appropriate variance, then $\Phi$ will satisfy RIP $\beta_s \leq \epsilon$ with probability $(1 - e^{-cM})$ if $M \geq cs \log(N/s)/\epsilon^2$, where $c$ and $C$ are constants depending on the distribution of the elements in $\Phi$.

Another random construction, which is often more useful in applications, is to use a sub-matrix of an orthogonal matrix. For example, let $\Phi$ be a (suitably normalised) sub-matrix of the matrix associated with the discrete Fourier transform, generated by drawing rows of the matrix at random (without replacement), then $\Phi$ will satisfy RIP $\beta_s \leq \epsilon$ with probability $(1 - e^{-cM})$ if $M \geq Cs \log^5 N \log(\epsilon^{-1})/\epsilon^2$, where $c$ and $C$ are constants.

## III. ITERATIVE HARD THRESHOLDING

### A. Definition of the Algorithm

In [15], we introduced the following Iterative Hard Thresholding algorithm ($IHT_s$). Let $\mathbf{y}^{[0]} = \mathbf{0}$ and use the iteration

$$\mathbf{y}^{[n+1]} = H_s(\mathbf{y}^{[n]} + \Phi^T(\mathbf{x} - \Phi \mathbf{y}^{[n]})), \tag{11}$$

where $H_s(\mathbf{a})$ is the non-linear operator that sets all but the largest (in magnitude) $s$ elements of $\mathbf{a}$ to zero. If there is no unique such set, a set can be selected either randomly or based on a predefined ordering of the elements. The convergence of this algorithm was proven in [15] under the condition that the operator norm $\|\Phi\|_2$ is smaller than one. In fact, the same argument can be used to show that the algorithm converges if our version of the restricted isometry property holds. The argument follows the same line as that in [15] and we omit the details here.

### B. Computational Complexity per Iteration

The iterative hard thresholding algorithm is very simple. It involves the application of the operators $\Phi$ and $\Phi^T$ once in each iteration as well as two vector additions. The operator $H_s$ involves a partial ordering of the elements of $\mathbf{a}^{[n]} = \mathbf{y}^{[n]} + \Phi^T(\mathbf{x} - \Phi \mathbf{y}^{[n]})$ in magnitude. The storage requirements are small. Apart from storage of $\mathbf{x}$, we only require the storage of the vector $\mathbf{a}$, which is of length $N$. Storage of $\mathbf{y}^{[n]}$, which has only $s$-non-zero elements, requires $2s$ numbers to be stored.

The computational bottle neck, both in terms of storage and computation time, is due to the operators $\Phi$ and $\Phi^T$. If these are general matrixes, the computational complexity and memory requirement is $O(MN)$. For large problems, it is common to use structured operators, basd for example on fast fourier transforms or Wavelet transforms, which require substantially less memory and can often be applied with $O(N \log M)$ or even $O(N)$ operations. In this case, the above algorithm has minimal computational requirements per iteration. If $\mathcal{L}$ is the complexity of applying the operators $\Phi$ and $\Phi^T$, then the computational complexity of the algorithm is $O(k^\star \mathcal{L})$, where $k^\star$ is the total number of iterations.

## IV. Iterative Hard Thresholding for Compressed Sensing

In this section we derive the main result of this paper. We show that if $\beta_{3s} < 1/8$, then the iterative hard thresholding algorithm reduces the estimation error in each iteration and is guaranteed to come within a constant factor of the best attainable estimation error. In fact, the algorithm needs a fixed number of iterations, depending only on the logarithm of a form of signal to noise ratio.

### A. Digest: The Main Result

The main result of this paper can be formally stated in the following theorem and corollary.

*Theorem 1:* Given a noisy observation $\mathbf{x} = \Phi\mathbf{y} + \mathbf{e}$, were $\mathbf{y}$ is an arbitrary vector. Let $\mathbf{y}^s$ be an approximation to $\mathbf{y}$ with no-more than s non-zero elements for which $\|\mathbf{y} - \mathbf{y}^s\|_2$ is minimal. If $\Phi$ has restricted isometry property with $\beta_{3s} < 1/8$, then, at iteration $k$, $IHT_s$ will recover an approximation $\mathbf{y}^k$ satisfying

$$\|\mathbf{y} - \mathbf{y}^k\|_2 \leq 2^{-k}\|\mathbf{y}^s\|_2 + 5\tilde{\epsilon}_s. \tag{12}$$

where

$$\tilde{\epsilon}_s = \|\mathbf{y} - \mathbf{y}^s\|_2 + \frac{1}{\sqrt{s}}\|\mathbf{y} - \mathbf{y}^s\|_1 + \|\mathbf{e}\|_2 \tag{13}$$

Furthermore, after at most

$$k^\star = \left\lceil \log_2\left(\frac{\|\mathbf{y}^s\|_2}{\tilde{\epsilon}_s}\right) \right\rceil \tag{14}$$

iterations, $IHT_s$ estimates $\mathbf{y}$ with accuracy

$$\|\mathbf{y} - \mathbf{y}^{k^\star}\|_2 \leq 6\left[\|\mathbf{y} - \mathbf{y}^s\|_2 + \frac{1}{\sqrt{s}}\|\mathbf{y} - \mathbf{y}^s\|_1 + \|\mathbf{e}\|_2\right]. \tag{15}$$

For exact sparse signals, we have a slightly better corollary.

*Corollary 1:* Given a noisy observation $\mathbf{x} = \Phi\mathbf{y}^s + \mathbf{e}$, were $\mathbf{y}^s$ is $s$-sparse. If $\Phi$ has the restricted isometry property with $\beta_{3s} < 1/8$, then, at iteration $k$, $IHT_s$ will recover an approximation $\mathbf{y}^k$ satisfying

$$\|\mathbf{y}^s - \mathbf{y}^k\|_2 \leq 2^{-k}\|\mathbf{y}^s\|_2 + 4\|\mathbf{e}\|_2. \tag{16}$$

Furthermore, after at most

$$k^\star = \left\lceil \log_2\left(\frac{\|\mathbf{y}^s\|_2}{\|\mathbf{e}\|_2}\right) \right\rceil \tag{17}$$

iterations, $IHT_s$ estimates $\mathbf{y}$ with accuracy

$$\|\mathbf{y}^s - \mathbf{y}^{k^\star}\|_2 \leq 5\|\mathbf{e}\|_2. \tag{18}$$

### B. Discussion of the Main Results

The main theorem states that the algorithm will find an approximation that comes close to the true vector. However, there is a limit to this. Asymptotically, we are only guaranteed to get as close as a multiple of

$$\tilde{\epsilon}_s = \|\mathbf{y} - \mathbf{y}^s\|_2 + \frac{1}{\sqrt{s}}\|\mathbf{y} - \mathbf{y}^s\|_1 + \|\mathbf{e}\|_2 \tag{19}$$

The quantity $\tilde{\epsilon}_s$ can be understood as an error term. This error term is composed of two components, the observation error $\mathbf{e}$ and the difference between the signal $\mathbf{y}$ and its best s term approximation $\mathbf{y}^s$. This makes intuitive sense. Assume, the observation error is zero and $\mathbf{y}$ is exactly $s$-sparse. In this case, the algorithm is guaranteed (under the conditions of the theorem) to find $\mathbf{y}$ exactly. For exact sparse signals, but with noisy observation, our success in recovering $\mathbf{y}$ is naturally limited by the size of the error. Assuming that $\mathbf{y}$ is not $s$-sparse, there will be an error between any $s$-term approximation and $\mathbf{y}$. The closest we can get to $\mathbf{y}$ with any $s$-sparse approximation is therefore limited by how well $\mathbf{y}$ can be approximated with $s$-sparse signals.

To show that this result is in fact optimal up to a constant, we can use the following argument. Theoretic considerations due to Kashin [17] and Garnaeva-Gluskin [18] show that any matrix $\Phi \in \mathbb{R}^{M \times N}$ must have at least $M \geq cs \log(N/s)$ for some constant $c$ in order for the observation $\mathbf{x} = \Phi\mathbf{y}$ to allow a reconstruction $\hat{\mathbf{y}}$ with $\|\mathbf{y} - \hat{\mathbf{y}}\|_2 \leq C/\sqrt{s}\|\mathbf{y}\|_1$. As discussed above, a random construction of $\Phi$ can achieve the RIP required in the

theorem with high probability if $M \geq cs \log(N/s)$. This shows that the dependence of the error guarantee of $IHT_s$ on $\frac{1}{\sqrt{s}}\|\mathbf{y} - \mathbf{y}^s\|_1$ is optimal up to a constant factor. Clearly, the dependence on $\|\mathbf{y} - \mathbf{y}^s\|_2$ is also unavoidable, as we cannot find any $s$-term approximation that beats the best $s$-term approximation, even if $\mathbf{y}$ would be known exactly. For a general bounded error $\mathbf{e}$, the dependence on $\|\mathbf{e}\|$ is also unavoidable, even if the support of $\mathbf{y}^s$ would be known, the error $\mathbf{e}$, projected onto this subspace, can induce an error depending on the size of $\mathbf{e}$. In summary, the error must depend on $\|\mathbf{y} - \mathbf{y}^s\|_2$ as there is no $s$-sparse approximation that could beat this. Furthermore, when projecting a signal into an $s$-dimensional observation space, the maximal information loss must lead to an error of the order $\frac{1}{\sqrt{s}}\|\mathbf{y} - \mathbf{y}^s\|_1$. Finally, the worst case estimation error must also depend on the size of the observation error $\mathbf{e}$.

The overall number of iterations required to achieve a desired accuracy depends on the logarithm of $\frac{\|\mathbf{y}^s\|_2}{\tilde{\epsilon}_s}$. We can think of the quantity $\frac{\|\mathbf{y}^s\|_2}{\tilde{\epsilon}_s}$ as a signal to noise ratio appropriate for sparse signal estimates. This term is large, whenever the observation noise is small and the signal $\mathbf{y}$ is well approximated by an $s$-sparse vector. Bounds on this term can be derived for a particular signal model, such as, for example, the *p-compressible* signal model for which we have presented the required bounds above.

### C. Derivation of the Error Bound

We now turn to the derivation of the main result. We start by proving the error bound in corollary 1. Let us recall some notation and introduce some abbreviations. We have

1) $\mathbf{x} = \Phi \mathbf{y}^s + \mathbf{e}$,
2) $\mathbf{r}^{[n]} = \mathbf{y}^s - \mathbf{y}^{[n]}$,
3) $\mathbf{a}^{[n+1]} = \mathbf{y}^{[n]} + \Phi^T(\mathbf{x} - \Phi \mathbf{y}^{[n]}) = \mathbf{y}^{[n]} + \Phi^T(\Phi \mathbf{y}^s + \mathbf{e} - \Phi \mathbf{y}^{[n]})$,
4) $\mathbf{y}^{[n+1]} = H_s(\mathbf{a}^{[n+1]})$, where $H_s$ is the hard thresholding operator that keeps the largest $s$ (in magnitude) elements and sets the other elements to zero.
5) $\Gamma^\star = \text{supp}\{\mathbf{y}^s\}$,
6) $\Gamma^n = \text{supp}\{\mathbf{y}^{[n]}\}$,
7) $B^{n+1} = \Gamma^\star \bigcup \Gamma^{n+1}$,

*Proof:* [Proof of the error bound in corollary 1] Consider the error

$$\|\mathbf{y}^s - \mathbf{y}^{[n+1]}\|_2 \leq \|\mathbf{y}^s_{B^{n+1}} - \mathbf{a}^{[n+1]}_{B^{n+1}}\|_2 + \|\mathbf{y}^{[n+1]}_{B^{n+1}} - \mathbf{a}^{[n+1]}_{B^{n+1}}\|_2. \tag{20}$$

Note that $\mathbf{y}^s - \mathbf{y}^{[n+1]}$ is supported on the set $B^{n+1} = \Gamma^\star \bigcup \Gamma^{n+1}$. By the thresholding operation, $\mathbf{y}^{[n+1]}$ is the best $s$-term approximation to $\mathbf{a}^{[n+1]}_{B^{n+1}}$. In particular, it is a better approximation than $\mathbf{y}^s$. This implies that $\|\mathbf{y}^{[n+1]} - \mathbf{a}^{[n+1]}_{B^{n+1}}\|_2 \leq \|\mathbf{y}^s - \mathbf{a}^{[n+1]}_{B^{n+1}}\|_2$ and we have

$$\|\mathbf{y}^s - \mathbf{y}^{[n+1]}\|_2 \leq 2\|\mathbf{y}^s_{B^{n+1}} - \mathbf{a}^{[n+1]}_{B^{n+1}}\|_2. \tag{21}$$

We now expand

$$\mathbf{a}^{[n+1]}_{B^{n+1}} = \mathbf{y}^{[n]}_{B^{n+1}} + \Phi^T_{B^{n+1}} \Phi \mathbf{r}^{[n]} + \Phi^T_{B^{n+1}} \mathbf{e}. \tag{22}$$

We then have

$$\begin{aligned}
\|\mathbf{y}^s - \mathbf{y}^{[n+1]}\|_2 &\leq 2\|\mathbf{y}^s_{B^{n+1}} - \mathbf{y}^{[n]}_{B^{n+1}} - \Phi^T_{B^{n+1}} \Phi \mathbf{r}^{[n]} - \Phi^T_{B^{n+1}} \mathbf{e}\|_2 \\
&\leq 2\|\mathbf{r}^{[n]}_{B^{n+1}} - \Phi^T_{B^{n+1}} \Phi \mathbf{r}^{[n]}\|_2 + 2\|\Phi^T_{B^{n+1}} \mathbf{e}\|_2 \\
&= 2\|(\mathbf{I} - \Phi^T_{B^{n+1}} \Phi_{B^{n+1}})\mathbf{r}^{[n]}_{B^{n+1}} - \Phi^T_{B^{n+1}} \Phi_{B^n \setminus B^{n+1}} \mathbf{r}^{[n]}_{B^n \setminus B^{n+1}}\|_2 \\
&\quad + 2\|\Phi^T_{B^{n+1}} \mathbf{e}\|_2 \\
&\leq 2\|(\mathbf{I} - \Phi^T_{B^{n+1}} \Phi_{B^{n+1}})\mathbf{r}^{[n]}_{B^{n+1}}\|_2 \\
&\quad + 2\|\Phi^T_{B^{n+1}} \Phi_{B^n \setminus B^{n+1}} \mathbf{r}^{[n]}_{B^n \setminus B^{n+1}}\|_2 \\
&\quad + 2\|\Phi^T_{B^{n+1}} \mathbf{e}\|_2
\end{aligned}$$

Now $B^n \setminus B^{n+1}$ is disjoint from $B^{n+1}$ and $|B^n \bigcup B^{n+1}| \leq 3s$. We can therefore use the bounds in (7), (10) and (9) and the fact that $\beta_{2s} \leq \beta_{3s}$

$$\|\mathbf{r}^{[n+1]}\|_2 \leq 2\beta_{2s}\|\mathbf{r}^{[n]}\|_2 + 2\beta_{3s}\|\mathbf{r}^{[n]}\|_2 + 2\|\mathbf{e}\|_2 \leq 4\beta_{3s}\|\mathbf{r}^{[n]}\|_2 + 2\|\mathbf{e}\|_2 \tag{23}$$

If $\beta_{3s} < \frac{1}{8}$, then

$$\|\mathbf{r}^{[n+1]}\|_2 < 0.5\|\mathbf{r}^{[n]}\|_2 + 2\|\mathbf{e}\|_2 \tag{24}$$

Iterating this relationship, and realising that $2(1 + 0.5 + 0.25 + \cdots) \leq 4$ and that $\mathbf{y}^{[0]} = \mathbf{0}$, we get

$$\|\mathbf{r}^{[k]}\|_2 < 2^{-k}\|\mathbf{y}^s\|_2 + 4\|\mathbf{e}\|_2 \tag{25}$$

This proves corollary 1. ∎

To prove the main theorem, we use the following two lemmas from [14]. The first lemma is stated here without proof, which follows that presented in [14], with the required correction for our definition of the RIP.

*Lemma 3 (Needell and Tropp, Proposition 3.5 in [14]):* Suppose the matrix $\Phi$ satisfies the RIP $\|\Phi\mathbf{y}^s\|_2 \leq \|\mathbf{y}^s\|_2$ for all $\mathbf{y}^s : \|\mathbf{y}^s\|_0 \leq s$, then for all vectors $\mathbf{y}$, the following bound holds

$$\|\Phi\mathbf{y}\|_2 \leq \|\mathbf{y}\|_2 + \frac{1}{\sqrt{s}}\|\mathbf{y}\|_1. \tag{26}$$

*Lemma 4 (Needell and Tropp, lemma 6.1 in [14]):* For any $\mathbf{y}$, let $\mathbf{y}^s$ be (the/any) best $s$-term approximation to $\mathbf{y}$. Let $\mathbf{y}_r = \mathbf{y} - \mathbf{y}^s$. Let

$$\mathbf{x} = \Phi\mathbf{y} + \mathbf{e} = \Phi\mathbf{y}^s + \Phi\mathbf{y}_r + \mathbf{e} = \Phi\mathbf{y}^s + \tilde{\mathbf{e}}. \tag{27}$$

If the RIP holds for sparsity $s$, then the norm of the error $\tilde{\mathbf{e}}$ can be bounded by

$$\|\tilde{\mathbf{e}}\|_2 \leq \|\mathbf{y} - \mathbf{y}^s\|_2 + \frac{1}{\sqrt{s}}\|\mathbf{y} - \mathbf{y}^s\|_1 + \|\mathbf{e}\|_2 \tag{28}$$

*Proof:*

$$\begin{aligned}
\|\tilde{\mathbf{e}}\|_2 &= \|\Phi\mathbf{y}_r + \mathbf{e}\|_2 = \|\Phi(\mathbf{y} - \mathbf{y}^s) + \mathbf{e}\|_2 \\
&\leq \|\Phi(\mathbf{y} - \mathbf{y}^s)\|_2 + \|\mathbf{e}\|_2 \\
&\leq \|(\mathbf{y} - \mathbf{y}^s)\|_2 + \frac{1}{\sqrt{s}}\|(\mathbf{y} - \mathbf{y}^s)\|_1 + \|\mathbf{e}\|_2,
\end{aligned}$$

where the last inequality follows form Lamma 3. ∎

*Proof:* [Proof of the error bound in theorem 1] To bound the error $\|\mathbf{y} - \mathbf{y}^{[k]}\|_2$, we note that

$$\begin{aligned}
\|\mathbf{y} - \mathbf{y}^{[k]}\|_2 &\leq \|\mathbf{r}^{[k]}\|_2 + \|\mathbf{y} - \mathbf{y}^s\|_2 \\
&\leq \|\mathbf{r}^{[k]}\|_2 + \|(\mathbf{y} - \mathbf{y}^s)\|_2 + \frac{1}{\sqrt{s}}\|(\mathbf{y} - \mathbf{y}^s)\|_1 + \|\mathbf{e}\|_2.
\end{aligned}$$

The proof of the main theorem then follows by bounding $\|\mathbf{r}^{[k]}\|_2$ using corollary 1 with $\tilde{\mathbf{e}}$ instead of $\mathbf{e}$ and lemma 4 to bound $\|\tilde{\mathbf{e}}\|_2$, that is

$$\|\mathbf{r}^{[k]}\|_2 \leq 2^{-k}\|\mathbf{y}^s\|_2 + 4\left[\|(\mathbf{y} - \mathbf{y}^s)\|_2 + \frac{1}{\sqrt{s}}\|(\mathbf{y} - \mathbf{y}^s)\|_1 + \|\mathbf{e}\|_2\right]. \tag{29}$$

∎

*D. Derivation of the Iteration Count*

*Proof:* [Proof of the second part of theorem 1] The first part of theorem 1 shows that

$$\|\mathbf{y} - \mathbf{y}^{[k]}\|_2 \leq 2^{-k}\|\mathbf{y}^s\|_2 + 5\tilde{\epsilon}_s, \tag{30}$$

where $\tilde{\epsilon}_s = \left[\|(\mathbf{y} - \mathbf{y}^s)\|_2 + \frac{1}{\sqrt{s}}\|(\mathbf{y} - \mathbf{y}^s)\|_1 + \|\mathbf{e}\|_2\right]$. We are therefore guaranteed to reduce the error to below any multiple $c$ of $\tilde{\epsilon}_s$, as long as $c > 5$. For example, assume we want to recover $\|\mathbf{y}\|$ with an error of less than $6\tilde{\epsilon}_s$. This implies that we require that

$$2^{-k}\|\mathbf{y}^s\|_2 \leq \tilde{\epsilon}_s \tag{31}$$

i.e. that

$$2^k \geq \frac{\|\mathbf{y}^s\|_2}{\tilde{\epsilon}_s}, \tag{32}$$

which in turn implies the second part of the theorem. The proof of the corresponding result in corollary 1 follows the same argument. ∎

## V. WHEN TO STOP

So far, we have given guarantees on the achievable error and a bound on the total number of iterations to achieve this bound. However, in practice, it is necessary to monitor quantities of the algorithm and decide to stop the iterations at some point. From the main result, it is clear that in general, we cannot do any better than to find an estimate with an error of $5\tilde{\epsilon}_s$. However, how do we know that we are getting close to this value?

A possible stopping criterion is $\|\mathbf{x} - \Phi\mathbf{y}^{[n]}\|_2 \leq \epsilon$. For this criterion, it is possible to use the same arguments as in appendix A of [14], to derive the following result.

*Lemma 5:* Assume that $\Phi$ satisfies the RIP with $\beta_{3s} < 1/8$. If at any iteration of $IHT_s$ the condition $\|\mathbf{x} - \Phi\mathbf{y}^{[n]}\|_2 \leq \epsilon$ holds, then

$$\|\mathbf{y} - \mathbf{y}^{[n]}\|_2 \leq 1.07(\epsilon + 2\tilde{\epsilon}_s), \tag{33}$$

where

$$\tilde{\epsilon}_s = \|\mathbf{y} - \mathbf{y}^s\|_2 + \frac{1}{\sqrt{s}}\|\mathbf{y} - \mathbf{y}^s\|_1 + \|\mathbf{e}\|_2. \tag{34}$$

Conversely, if at any iteration of $IHT_s$ the condition

$$\|\mathbf{y} - \mathbf{y}^{[n]}\|_2 \leq \epsilon - 1/\sqrt{s}\|\mathbf{y} - \mathbf{y}^s\|_1 - \|\mathbf{e}\|_2 \tag{35}$$

holds, then $\|\mathbf{x} - \Phi\mathbf{y}^{[n]}\|_2 \leq \epsilon$.

This lemma can be used to calculate a stopping criterion for $IHT_s$. For example, if we want to estimate $\|\mathbf{y}\|_2$ with accuracy $c\tilde{\epsilon}_s$, we know that we are done as soon as $\|\mathbf{x} - \Phi\mathbf{y}_2^{[n]}\|_2 \leq (c/1.07 - 2)\tilde{\epsilon}_s$. Note that in general, $IHT_s$ is only guaranteed to work for $c > 5$, however, as soon as we observe that $\|\mathbf{x} - \Phi\mathbf{y}^{[n]}\|_2 \leq (c/1.07 - 2)$, we know that the estimation error must be below $c\tilde{\epsilon}_s$.

*Proof:* To prove the first part, note that the stopping criterion implies that

$$\begin{aligned} \epsilon \geq \|\Phi(\mathbf{y} - \mathbf{y}^{[n]}) + \mathbf{e}\|_2 &= \|\Phi(\mathbf{y}^s - \mathbf{y}^{[n]}) + \tilde{\mathbf{e}}\|_2 \\ &\geq \sqrt{1 - \beta_{2s}}\|\mathbf{y}^s - \mathbf{y}^{[n]}\|_2 - \|\tilde{\mathbf{e}}\|_2, \end{aligned}$$

so that

$$\|\mathbf{y}^s - \mathbf{y}^{[n]}\|_2 \leq \frac{\epsilon + \|\tilde{\mathbf{e}}\|_2}{\sqrt{1 - \beta_{2s}}}. \tag{36}$$

Furthermore,

$$\|\mathbf{y} - \mathbf{y}^{[n]}\|_2 \leq \|\mathbf{y}^s - \mathbf{y}^{[n]}\|_2 + \|\mathbf{y} - \mathbf{y}^s\|_2, \tag{37}$$

so that (using the bound on $\|\tilde{\mathbf{e}}\|_2$ from lemma 4)

$$\begin{aligned} \|\mathbf{y} - \mathbf{y}^{[n]}\|_2 &\leq \frac{\epsilon + \|\tilde{\mathbf{e}}\|_2}{\sqrt{1 - \beta_{2s}}} + \|\mathbf{y} - \mathbf{y}^s\|_2 \\ &\leq \frac{\epsilon + 2\|\mathbf{y} - \mathbf{y}^s\|_2 + \frac{1}{\sqrt{s}}\|\mathbf{y} - \mathbf{y}^s\|_1 + \|\mathbf{e}\|_2}{\sqrt{1 - \beta_{2s}}}. \\ &\leq \frac{\epsilon + 2\tilde{\epsilon}_s}{\sqrt{1 - \beta_{2s}}}. \end{aligned}$$

This proves the first part of the lemma using $\beta_{2s} \leq \beta_{3s} < 1/8$.

To prove the second part, note that if

$$\|\mathbf{y} - \mathbf{y}^{[n]}\|_2 \leq \epsilon - 1/\sqrt{s}\|\mathbf{y} - \mathbf{y}^{[n]}\|_1 - \|\mathbf{e}\|_2. \tag{38}$$

holds, then

$$\begin{aligned} \epsilon &\geq \|\mathbf{y} - \mathbf{y}^{[n]}\|_2 + 1/\sqrt{s}\|\mathbf{y} - \mathbf{y}^{[n]}\|_1 + \|\mathbf{e}\|_2 \\ &\geq \|\Phi(\mathbf{y} - \mathbf{y}^{[n]})\|_2 + \|\mathbf{e}\|_2 \\ &\geq \|\Phi(\mathbf{y} - \mathbf{y}^{[n]}) + \mathbf{e}\|_2. \end{aligned}$$

We have here used lemma 3 to bound $\|\Phi(\mathbf{y} - \mathbf{y}^{[n]})\|_2 \leq \|(\mathbf{y} - \mathbf{y}^{[n]})\|_2 + \frac{1}{\sqrt{s}}\|(\mathbf{y} - \mathbf{y}^{[n]})\|_1$, where $s$ is such that RIP holds for this $s$.

∎

## VI. Comparison to CoSaMP

In [14] the authors introduced a subspace pursuit algorithm called CoSaMP, which offers similar guarantees to the iterative hard thresholding approach of this paper. The result for CoSaMP is as follows

*Theorem 2 (Needell & Tropp [14]):* If $\Phi$ has the (normal) restricted isometry property with $\delta_{4s} \leq 0.1$, then, at iteration $k$, CoSaMP will recover an approximation $\mathbf{y}^k$ satisfying

$$\|\mathbf{y}^s - \mathbf{y}^k\|_2 \leq 2^{-k}\|\mathbf{y}^s\|_2 + 15\|\mathbf{e}\|_2 \tag{39}$$

if $\mathbf{x} = \Phi\mathbf{y}^s + \mathbf{e}$ for $\mathbf{y}^s$ $s$-sparse and

$$\|\mathbf{y} - \mathbf{y}^k\|_2 \leq 2^{-k}\|\mathbf{y}\|_2 + 20\tilde{\epsilon}_s, \tag{40}$$

if $\mathbf{x} = \Phi\mathbf{y} + \mathbf{e}$ for all $\mathbf{y}$.

Two remarks are in order. Firstly, for $IHT_s$, we require $\beta_{3s} \leq 0.125$, whilst for CoSaMP, $\delta_{4s} \leq 0.1$ is required. Now, $\beta$ and $\delta$ are related as follows

$$\frac{\beta}{2-\beta} = \delta, \tag{41}$$

therefore, $IHT_s$ requires $\delta_{3s} \leq 0.0667$. To compare this to the requirement for CoSaMP, we use corollary 3.4 from [14], which states that for integers $a$ and $s$, $\delta_{as} \leq a\delta_{2s}$. Therefore, if $\delta_{2s} \leq 0.025$, the condition for CoSaMP is satisfied, whilst for $IHT_s$, we have a comparable condition requiring that $\delta_{2s} \leq 0.0222$.

Whilst the requirement on $\delta_{2s}$ is marginally weaker for CoSaMP as compared to $IHT_s$, $IHT_s$ is guaranteed to get a roughly four times lower approximation error. For example, in the exact sparse case, $IHT_s$ is guaranteed to calculate an error approaching $4\|\mathbf{e}\|_2$, which should be compared to the guarantee of $15\|\mathbf{e}\|_2$ for CoSaMP. For general signals, the guarantees are $5\tilde{\epsilon}_s$ for $IHT_s$ and $20\tilde{\epsilon}_s$ for CoSaMP.

The number of iterations required for $IHT_s$ is logarithmical in the signal to noise ratio. This means that for noiseless observations, $IHT_s$ would require an infinite number of iterations to reduce the error to zero. This is a well known property of algorithms that use updates of the form $\mathbf{y}^{[n]} + \Phi^T(\mathbf{x} - \Phi\mathbf{y}^{[n]})$. CoSaMP on the other hand is guaranteed to estimate $\mathbf{y}$ with precision $20\tilde{\epsilon}_s$ in at most $6(s+1)$ iterations, however, to achieve this, CoSaMP requires the solution to an inverse problem in each iteration, which is costly. $IHT_s$ does not require the exact solution to an inverse problem. If CoSaMP is implemented using fast partial solutions to the inverse problems, the iteration count guarantees become similar to the ones derived here for $IHT_s$.

## VII. What's in a Theorem

A word of caution is in order. We have here shown that the Iterative Hard Thresholding algorithm has theoretical properties, which are comparable to those of other state of the art algorithms such as CoSaMP and has recovery guarantees of the same order as $\ell_1$ based approaches.

At a first glance, this seems to be at odds with previously reported numerical results [15], which have shown that $IHT_s$ does not perform as well as other methods such as Orthogonal Matching Pursuit, for which there are currently no comparable performance guarantees. What is more, $IHT_s$ does also perform less well in numerical studies than $CoSaMP$ or $\ell_1$ based approaches.

To understand this disparity, it has to be realised that the uniform performance guarantees derived here for $IHT_s$ and elsewhere for CoSaMP and $\ell_1$ based methods are worst case bounds, that is, they guarantee the performance of the algorithms in the worst possible scenario. Numerical studies on the other hand cannot in general test this worst case behaviour. This is because we do not in general know what particular signal would be the most difficult to recovered. Numerical experiments therefore analyse average behaviour of the methods, that is, they study the recovery of *typical* signals.

Whilst uniform guarantees can be derived for the $IHT_s$ algorithm, the difference in numerically observed performance between this method and other algorithms with similar uniform recovery guarantees indicates that uniform guarantees are not necessarily a good measure to indicate good average performance. This is further confirmed when looking at the average case analysis of algorithms such as Orthogonal Matching Pursuit [9], for which the uniform guarantees are currently available are markedly different to those of $IHT_s$, CoSaMP and $\ell_1$ methods [10], whilst in numerical studies, OMP can under certain conditions outperform some of these approaches.

## VIII. CONCLUSION

The abstract made eight claims regarding the performance of the iterative hard thresholding algorithm. Let us here summarise these in somewhat more detail.

- Error guarantee; We have shown that an estimation error of $6\|\tilde{\mathbf{e}}\|_2$ can be achieved within a finite number of iterations.
- Robustness to noise; We have shown that the achievable estimation error depends linearly on the size of the observation error. Performance therefore degrades linearly if the noise is increased.
- Minimum number of observations; The requirement on the isometry constant dictates that the number of observations grows linearly with $s$ and logarithmically with $N$. Up to a constant, this relation is known to be the best attainable.
- Sampling operator; The algorithm is simple and requires only the application of $\Phi$ and $\Phi^T$.
- Memory requirement; We have shown this to be linear in the problem size, if we can ignore the storage requirement for $\Phi$ and $\Phi^T$.
- Computational complexity; We could shown that the computational complexity is of the same order as the application of the measurement operator or its adjoint per iteration. The total number of iterations is bounded due to the linear convergence of the algorithm and depends logarithmically on the signal to noise ratio $\|\mathbf{y}^s\|_2/\|\tilde{\mathbf{e}}\|_2$.
- Number of iterations. We have shown that after at most $\left\lceil \log_2\left(\frac{\|\mathbf{y}^s\|_2}{\tilde{\epsilon}_s}\right) \right\rceil$ iterations, the estimation error is smaller than $6\tilde{\epsilon}_s$.
- Uniform performance guarantees. The results presented here only depend on $\beta_{3s}$ and do not depend on the size and distribution of the largest $s$ elements in $\mathbf{y}$.

This is an impressive list of properties for such a relatively simple algorithm. To our knowledge, only the CoSaMP algorithm shares similar guarantees. However, as discussed in section VII, uniform guarantees are not the only consideration and in practice marked differences in the average performance of different methods are apparent. For many small problems, the restricted isometry property of random matrices is often too large to explain the behaviour of the different methods in these studies. Furthermore, it has long been observed, that the distribution of the magnitude of the non-zero coefficients also has an important influence on the performance of different methods. Whilst the theoretical guarantees derived in this and similar papers are important to understand the behaviour of an algorithm, it is also clear that other facts have to be taken into account in order to predict the *typical* performance of algorithms in many practical situations.

## ACKNOWLEDGMENTS

This research was supported by EPSRC grant D000246/1. MED acknowledges support of his position from the Scottish Funding Council and their support of the Joint Research Institute with the Heriot-Watt University as a component part of the Edinburgh Research Partnership.

## REFERENCES

[1] H. Nyquist, "Certain topics in telegraph transmission theory," *Transactions of the A. I. E. E.*, pp. 617–644, Feb. 1928.
[2] C. A. Shannon and W. Weaver, *The mathematical theory of communication*. University of Illinois Press, 1949.
[3] I. F. Gorodnitsky, J. S. George, and B. D. Rao, "Neuromagnetic source imaging with focuss: a recursive weighted minimum norm algorithm," *Neurophysiology*, vol. 95, no. 4, pp. 231–251, 1995.
[4] E. Candès, J. Romberg, and T. Tao, "Robust uncertainty principles: Exact signal reconstruction from highly incomplete frequency information." *IEEE Transactions on information theory*, vol. 52, pp. 489–509, Feb 2006.
[5] E. Candès, J. Romberg, and T. Tao, "Stable signal recovery from incomplete and inaccurate measurements," *Comm. Pure Appl. Math.*, vol. 59, no. 8, pp. 1207–1223, 2006.
[6] E. Candès and J. Romberg, "Quantitative robust uncertainty principles and optimally sparse decompositions," *Foundations of Comput. Math*, vol. 6, no. 2, pp. 227 – 254, 2006.
[7] D. Donoho, "Compressed sensing," *IEEE Trans. on Information Theory*, vol. 52, no. 4, pp. 1289–1306, 2006.
[8] M. Vetterli, P. Marziliano, and T. Blu, "Sampling signals with finite rate of innovation," *IEEE Transactions on Signal Processing*, vol. 50, no. 6, pp. 1417–1428, 2002.
[9] S. Mallat, G. Davis, and Z. Zhang, "Adaptive time-frequency decompositions," *SPIE Journal of Optical Engineering*, vol. 33, pp. 2183–2191, July 1994.
[10] J. A. Tropp and A. C. Gilbert, "Signal recovery from partial information via orthogonal matching pursuit," *Submitted for publication,*, 2006.

[11] D. Needell and R. Vershynin, "Uniform uncertainty principle and signal recovery via regularized orthogonal matching pursuit," *submitted*, 2007.
[12] D. Needell and R. Vershynin, "Signal recovery from incomplete and inacurate measurements via regularized orthogonal matching pursuit.," *submitted*, 2008.
[13] W. Dai and O. Milenkovic, "Subspace pursuit for compressed sensing: Closing the gap between performance and complexity," *submitted*, 2008.
[14] D. Needell and J. Tropp, "COSAMP: Iterative signal reovery from incomplete and inacurate samples.," *submitted*, 2008.
[15] T. Blumensath and M. Davies, "Iterative thresholding for sparse approximations," *to appear in Journal of Fourier Analysis and Applications, special issue on sparsity*, 2008.
[16] R. A. Horn and C. R. Johnson, *Matrix Analysis*. Cambridge University Press, 1985.
[17] B. Kashin, "The width of certain finite dimensional sets and classes of smooth functions," *Izvestia*, vol. 41, pp. 334–351.
[18] A. Garnaev and E. Gluskin, "On width of the Euclidean ball," *Sov. Math. Dokl*, 1984.